\documentclass[twoside,11pt]{article}

\usepackage{jmlr2e}

\ShortHeadings{Learning with Mixtures of Trees}{Meil\u{a} and Jordan}
\firstpageno{1}

\usepackage{natbib}

\usepackage{glossaries}

\def\N{{\mathbb{N}}}  \def\R{{\mathbb{R}}} \def\C{{\mathbb{C}}}   
\def\Eb{{\mathbb{E}}}

\newcommand{\beq}{\begin{eqnarray}}
\newcommand{\eeq}{\end{eqnarray}}

\def\cA{{\mathcal{A}}}  \def\cC{{\mathcal{C}}}

 \def\cN{{\mathcal{N}}}

\newacronym{MLE}{MLE}{maximum likelihood estimate}
\newacronym{BLUE}{BLUE}{best linear unbiased estimator}
\newacronym{WLS}{WLS}{weighted least squares}

\usepackage{amsmath,amssymb,amsfonts,mathrsfs,bm}
\usepackage{mathtools}
\usepackage{amsthm}
\usepackage{scalerel}
\usepackage{nicefrac}
\usepackage[shortlabels]{enumitem}
\usepackage{graphicx}
\usepackage{epstopdf}
\DeclareGraphicsExtensions{.eps,.png,.jpg,.pdf}

\usepackage{url}
\usepackage{colortbl}
\usepackage{booktabs}
\usepackage{multirow}
\usepackage[table,dvipsnames]{xcolor}
\usepackage[normalem]{ulem}
\usepackage{xparse}
\usepackage{calc}
\usepackage{etoolbox}

\makeatletter
\@ifpackageloaded{natbib}{
	\relax
}{
	\usepackage{cite}
}
\makeatother

\usepackage{array}
\newcolumntype{L}[1]{>{\raggedright\let\newline\\\arraybackslash\hspace{0pt}}m{#1}}
\newcolumntype{C}[1]{>{\centering\let\newline\\\arraybackslash\hspace{0pt}}m{#1}}
\newcolumntype{R}[1]{>{\raggedleft\let\newline\\\arraybackslash\hspace{0pt}}m{#1}}

\makeatletter
\let\MYcaption\@makecaption
\makeatother
\usepackage[font=footnotesize]{subcaption}
\makeatletter
\let\@makecaption\MYcaption
\makeatother

\usepackage{glossaries}
\makeatletter
\let\oldgls\gls
\let\oldglspl\glspl

\newcommand\fussy@ifnextchar[3]{%
	\let\reserved@d=#1%
	\def\reserved@a{#2}%
	\def\reserved@b{#3}%
	\futurelet\@let@token\fussy@ifnch}
\def\fussy@ifnch{%
	\ifx\@let@token\reserved@d
		\let\reserved@c\reserved@a
	\else
		\let\reserved@c\reserved@b
	\fi
	\reserved@c}

\renewcommand{\gls}[1]{%
\oldgls{#1}\fussy@ifnextchar.{\@checkperiod}{\@}}
\renewcommand{\glspl}[1]{%
\oldglspl{#1}\fussy@ifnextchar.{\@checkperiod}{\@}}

\newcommand{\@checkperiod}[1]{%
	\ifnum\sfcode`\.=\spacefactor\else#1\fi
}
\makeatother

\newacronym{wrt}{w.r.t.}{with respect to}
\newacronym{RHS}{R.H.S.}{right-hand side}
\newacronym{LHS}{L.H.S.}{left-hand side}
\newacronym{iid}{i.i.d.}{independent and identically distributed}
\newacronym{SVD}{SVD}{singular value decomposition}
\newacronym{5G}{5G}{fifth generation wireless}
\newacronym{3GPP}{3GPP}{third generation partnership project}
\newacronym{OFDM}{OFDM}{orthogonal frequency-division multiplexing}

\usepackage{float}

\usepackage[capitalize]{cleveref}
\crefname{equation}{}{}
\Crefname{equation}{}{}
\crefname{claim}{claim}{claims}
\crefname{step}{step}{steps}
\crefname{line}{line}{lines}
\crefname{condition}{condition}{conditions}
\crefname{dmath}{}{}
\crefname{dseries}{}{}
\crefname{dgroup}{}{}

\crefname{Problem}{Problem}{Problems}
\crefformat{Problem}{Problem~(#2#1#3)}
\crefrangeformat{Problem}{Problems~(#3#1#4) to~(#5#2#6)}

\crefname{Theorem}{Theorem}{Theorems}
\crefname{Corollary}{Corollary}{Corollaries}
\crefname{Proposition}{Proposition}{Propositions}
\crefname{Lemma}{Lemma}{Lemmas}
\crefname{Definition}{Definition}{Definitions}
\crefname{Example}{Example}{Examples}
\crefname{Assumption}{Assumption}{Assumptions}
\crefname{Remark}{Remark}{Remarks}
\crefname{Rem}{Remark}{Remarks}
\crefname{remarks}{Remarks}{Remarks}
\crefname{Appendix}{Appendix}{Appendices}
\crefname{Supplement}{Supplement}{Supplements}
\crefname{Exercise}{Exercise}{Exercises}
\crefname{Theorem_A}{Theorem}{Theorems}
\crefname{Corollary_A}{Corollary}{Corollaries}
\crefname{Proposition_A}{Proposition}{Propositions}
\crefname{Lemma_A}{Lemma}{Lemmas}
\crefname{Definition_A}{Definition}{Definitions}

\usepackage{crossreftools}
\ifx\notloadhyperref\undefined
\else
	\relax
\fi

\usepackage{algorithm,algorithmic}

\ifx\loadbreqn\undefined
	\relax
\else
	\usepackage{breqn}
\fi

\makeatletter
\def\cleartheorem#1{%
    \expandafter\let\csname#1\endcsname\relax
    \expandafter\let\csname c@#1\endcsname\relax
}
\def\clearthms#1{ \@for\tname:=#1\do{\cleartheorem\tname} }
\makeatother

\ifx\renewtheorem\undefined
	\ifx\useTheoremCounter\undefined
		\newtheorem{Theorem}{Theorem}
		\newtheorem{Corollary}{Corollary}
		\newtheorem{Proposition}{Proposition}
		\newtheorem{Lemma}{Lemma}
	\else
		\newtheorem{Theorem}{Theorem}
		\newtheorem{Corollary}[Theorem]{Corollary}
		\newtheorem{Proposition}[Theorem]{Proposition}
	\fi

	\newtheorem{Definition}{Definition}
	
	\newtheorem{Remark}{Remark}

\fi

\theoremstyle{remark}

\theoremstyle{plain}

\newcommand{\qednew}{\nobreak \ifvmode \relax \else
		\ifdim\lastskip<1.5em \hskip-\lastskip
			\hskip1.5em plus0em minus0.5em \fi \nobreak
		\vrule height0.75em width0.5em depth0.25em\fi}



\NewDocumentCommand{\movedownsub}{e{^_}}{%
	\IfNoValueTF{#1}{%
		\IfNoValueF{#2}{^{}}
	}{%
		^{#1}
	}%
	\IfNoValueF{#2}{_{#2}}
}

\let\latexchi\chi
\RenewDocumentCommand{\chi}{}{\latexchi\movedownsub}

\newcommand{\calN}{\mathcal{N}}

\newcommand{\bA}{\mathbf{A}}

\newcommand{\bu}{\mathbf{u}}

\newcommand{\bv}{\mathbf{v}}

\newcommand{\bx}{\mathbf{x}}
\newcommand{\bX}{\mathbf{X}}
\newcommand{\by}{\mathbf{y}}

\newcommand{\bz}{\mathbf{z}}

\DeclareSymbolFont{bsfletters}{OT1}{cmss}{bx}{n}
\DeclareSymbolFont{ssfletters}{OT1}{cmss}{m}{n}
\DeclareMathSymbol{\bsfGamma}{0}{bsfletters}{'000}
\DeclareMathSymbol{\ssfGamma}{0}{ssfletters}{'000}
\DeclareMathSymbol{\bsfDelta}{0}{bsfletters}{'001}
\DeclareMathSymbol{\ssfDelta}{0}{ssfletters}{'001}
\DeclareMathSymbol{\bsfTheta}{0}{bsfletters}{'002}
\DeclareMathSymbol{\ssfTheta}{0}{ssfletters}{'002}
\DeclareMathSymbol{\bsfLambda}{0}{bsfletters}{'003}
\DeclareMathSymbol{\ssfLambda}{0}{ssfletters}{'003}
\DeclareMathSymbol{\bsfXi}{0}{bsfletters}{'004}
\DeclareMathSymbol{\ssfXi}{0}{ssfletters}{'004}
\DeclareMathSymbol{\bsfPi}{0}{bsfletters}{'005}
\DeclareMathSymbol{\ssfPi}{0}{ssfletters}{'005}
\DeclareMathSymbol{\bsfSigma}{0}{bsfletters}{'006}
\DeclareMathSymbol{\ssfSigma}{0}{ssfletters}{'006}
\DeclareMathSymbol{\bsfUpsilon}{0}{bsfletters}{'007}
\DeclareMathSymbol{\ssfUpsilon}{0}{ssfletters}{'007}
\DeclareMathSymbol{\bsfPhi}{0}{bsfletters}{'010}
\DeclareMathSymbol{\ssfPhi}{0}{ssfletters}{'010}
\DeclareMathSymbol{\bsfPsi}{0}{bsfletters}{'011}
\DeclareMathSymbol{\ssfPsi}{0}{ssfletters}{'011}
\DeclareMathSymbol{\bsfOmega}{0}{bsfletters}{'012}
\DeclareMathSymbol{\ssfOmega}{0}{ssfletters}{'012}

\makeatletter
\newcommand*\rel@kern[1]{\kern#1\dimexpr\macc@kerna}
\newcommand*\widebar[1]{%
  \begingroup
  \def\mathaccent##1##2{%
    \rel@kern{0.8}%
    \overline{\rel@kern{-0.8}\macc@nucleus\rel@kern{0.2}}%
    \rel@kern{-0.2}%
  }%
  \macc@depth\@ne
  \let\math@bgroup\@empty \let\math@egroup\macc@set@skewchar
  \mathsurround\z@ \frozen@everymath{\mathgroup\macc@group\relax}%
  \macc@set@skewchar\relax
  \let\mathaccentV\macc@nested@a
  \macc@nested@a\relax111{#1}%
  \endgroup
}
\makeatother

\DeclareMathOperator*{\argmax}{arg\,max}

\DeclareMathOperator{\tr}{tr}

\newcommand{\ifbcdot}[1]{\ifblank{#1}{\cdot}{#1}}

\DeclarePairedDelimiterX\abs[1]{\lvert}{\rvert}{\ifbcdot{#1}}
\DeclarePairedDelimiterX\parens[1]{(}{)}{\ifbcdot{#1}}
\DeclarePairedDelimiterX\brk[1]{[}{]}{\ifbcdot{#1}}
\DeclarePairedDelimiterX\braces[1]{\{}{\}}{\ifbcdot{#1}}
\DeclarePairedDelimiterX\angles[1]{\langle}{\rangle}{\ifblank{#1}{\cdot,\cdot}{#1}}
\DeclarePairedDelimiterX\ip[2]{\langle}{\rangle}{\ifbcdot{#1},\ifbcdot{#2}}
\DeclarePairedDelimiterX\norm[1]{\lVert}{\rVert}{\ifbcdot{#1}}
\DeclarePairedDelimiterX\ceil[1]{\lceil}{\rceil}{\ifbcdot{#1}}
\DeclarePairedDelimiterX\floor[1]{\lfloor}{\rfloor}{\ifbcdot{#1}}

\DeclarePairedDelimiterXPP\trace[1]{\operatorname{Tr}}{(}{)}{}{\ifbcdot{#1}} 
\DeclarePairedDelimiterXPP\col[1]{\operatorname{col}}{\{}{\}}{}{\ifbcdot{#1}} 
\DeclarePairedDelimiterXPP\row[1]{\operatorname{row}}{\{}{\}}{}{\ifbcdot{#1}} 
\DeclarePairedDelimiterXPP\erf[1]{\operatorname{erf}}{(}{)}{}{\ifbcdot{#1}}
\DeclarePairedDelimiterXPP\erfc[1]{\operatorname{erfc}}{(}{)}{}{\ifbcdot{#1}}
\DeclarePairedDelimiterXPP\KLD[2]{D}{(}{)}{}{\ifbcdot{#1}\, \delimsize\|\, \ifbcdot{#2}} 
\DeclarePairedDelimiterXPP\op[2]{\operatorname{#1}}{(}{)}{}{#2} 

\newcommand{\bone}{\bm{1}}

\DeclarePairedDelimiterXPP\indicate[1]{{\bf 1}}{\{}{\}}{}{\ifbcdot{#1}}

\providecommand\given{}

\DeclarePairedDelimiterX\Set[2]\{\}{%
\renewcommand\given{\SetSymbol[\delimsize]{#1}}
#2
}
\DeclarePairedDelimiterX\Setc[1]\{\}{%
\renewcommand\given{\SetSymbol{:}}
#1
}

\NewDocumentCommand\set{s o m}{%
	\IfBooleanTF#1%
	{\IfValueTF{#2}{\Set*{#2}{#3}}{\Setc*{#3}}}%
	{\IfValueTF{#2}{\Set{#2}{#3}}{\Setc{#3}}}%
}

\NewDocumentCommand{\evalat}{ s O{\big} m e{_^} }{%
\IfBooleanTF{#1}%
{\left. #3 \right|}{#3#2|}%
\IfValueT{#4}{_{#4}}%
\IfValueT{#5}{^{#5}}%
}

\NewDocumentCommand \ifcondp {m m} {%
	#1%
	\IfValueT{#2}{\given #2}%
}

\providecommand\given{}
\DeclarePairedDelimiterXPP\cprob[1]{}(){}{
\renewcommand\given{\nonscript\,\delimsize\vert\allowbreak\nonscript\,\mathopen{}}
\ifcondp#1
}
\DeclarePairedDelimiterXPP\cexp[1]{}[]{}{
\renewcommand\given{\nonscript\,\delimsize\vert\allowbreak\nonscript\,\mathopen{}}
\ifcondp#1
}

\DeclareDocumentCommand \P { s e{_^} >{\SplitArgument{ 1 }{ @| }}d() g } {%
	\mathbb{P}%
	\IfBooleanTF{#1}%
		{
			\IfValueT{#2}{_{#2}}%
			\IfValueT{#3}{^{#3}}%
			\IfValueTF{#5}%
				{\cprob{#4 \given #5}}%
				{\IfValueT{#4}{\cprob{#4}}}%
		}%
		{
			\IfValueT{#2}{_{#2}}%
			\IfValueT{#3}{^{#3}}%
			\IfValueTF{#5}%
				{\cprob*{#4 \given #5}}%
				{\IfValueT{#4}{\cprob*{#4}}}%
		}%
}

\DeclareDocumentCommand \E { s e{_^} >{\SplitArgument{ 1 }{ @| }}d[] g } {%
	\mathbb{E}%
	\IfBooleanTF{#1}%
		{
			\IfValueT{#2}{_{#2}}%
			\IfValueT{#3}{^{#3}}%
			\IfValueTF{#5}%
				{\cexp{#4 \given #5}}%
				{\IfValueT{#4}{\cexp{#4}}}%
		}%
		{
			\IfValueT{#2}{_{#2}}%
			\IfValueT{#3}{^{#3}}%
			\IfValueTF{#5}%
				{\cexp*{#4 \given #5}}%
				{\IfValueT{#4}{\cexp*{#4}}}%
		}%
}

\ExplSyntaxOn
\NewDocumentCommand \dist {m o o} {%
\mathrm{#1}\left(%
	\IfValueT{#3}{%
		\tl_if_blank:nTF{ #3 }{\cdot\, \middle|\, }{#3\, \middle|\, }%
	}
	\IfValueT{#2}{#2}%
\right)%
}
\ExplSyntaxOff

\ifdefined\N
    \renewcommand{\N}[2]{\dist{\calN}[#1,\, #2]}
\else
    \newcommand{\N}[2]{\dist{\calN}[#1,\, #2]}
\fi

\NewDocumentCommand {\cbrace} {t+ D[]{black} D(){\widthof{#5}} m m } {%
	\begingroup%
		\color{#2}
		\IfBooleanTF{#1}{%
			\overbrace{#4}^%
		}{
			\underbrace{#4}_%
		}%
		{\parbox[c]{#3}{\centering\footnotesize{#5}}}%
	\endgroup%
}

\let\oldforall\forall
\renewcommand{\forall}{\oldforall \, }

\let\oldexist\exists
\renewcommand{\exists}{\oldexist \, }

\graphicspath{{./Figures/}{./figures/}}
\DeclareDocumentCommand{\includeCroppedPdf}{ o O{./Figures/} m }{
	\IfFileExists{#2#3-crop.pdf}{}{%
		\immediate\write18{pdfcrop #2#3.pdf #2#3-crop.pdf}}%
	\includegraphics[#1]{#2#3-crop.pdf}
}

\makeatletter
\newcommand*{\addFileDependency}[1]{
  \typeout{(#1)}
  \@addtofilelist{#1}
  \IfFileExists{#1}{}{\typeout{No file #1.}}
}
\makeatother

\definecolor{gray90}{gray}{0.9}

\ifx\nohighlights\undefined

	\newcommand{\msout}[1]{\text{\color{green} \sout{\ensuremath{#1}}}}
	\newcommand{\del}[1]{{\color{green}\ifmmode \msout{#1}\else\sout{#1}\fi}}
\else

	\newcommand{\msout}[1]{#1}
	\newcommand{\del}[1]{#1}
\fi

\newcommand{\hhide}[1]{}

\ifx\diagnoselabel\undefined
	\relax
\else
	\makeatletter
	\def\@testdef #1#2#3{%
		\def\reserved@a{#3}\expandafter \ifx \csname #1@#2\endcsname
			\reserved@a  \else
			\typeout{^^Jlabel #2 changed:^^J%
				\meaning\reserved@a^^J%
				\expandafter\meaning\csname #1@#2\endcsname^^J}%
			\@tempswatrue \fi}
	\makeatother
\fi

\usepackage{fancyhdr}
\begin{document}

\title{Restricted Isometry Property of Rank-One Measurements with Random Unit-Modulus Vectors}

\author{\name Wei Zhang \email zhangwei.sz@hit.edu.cn \\
       \addr School of Electronics and Information Engineering\\
       Harbin Institute of Technology Shenzhen\\
       Shenzhen, 518000, China
       \AND
       \name Zhenni Wang \email zhenni126@126.com \\
       \addr Department of Electrical Engineering\\
       City University of Hong Kong\\
       Hong Kong, 999077, Hong Kong China}

\editor{}

\maketitle

\begin{abstract}
The restricted isometry property (RIP) is essential for the linear map to guarantee the successful recovery of low-rank matrices. 
The existing works show that the linear map generated by the measurement matrices  with \gls{iid} entries satisfies RIP with high probability. 
However, when dealing with non-\gls{iid} measurement matrices, such as the rank-one measurements,  the RIP compliance may not be guaranteed.
In this paper, we show that the RIP can still be achieved with high probability, when the rank-one measurement matrix is constructed by the random unit-modulus vectors.
Compared to the existing works, we first address the challenge of establishing RIP for the linear map in non-\gls{iid} scenarios.
As validated in the experiments, this linear map is memory-efficient, and not only satisfies the RIP but also exhibits similar recovery performance of the low-rank matrices to that of conventional \gls{iid} measurement matrices.
\end{abstract}

\section{Introduction}
The low-rank matrix recovery is a popular topic in many fields, such as wireless communication, signal processing, and image processing \citep{candes2013phaselift,chen2015exact,shechtman2015phase,davenport2016overview,zhang2018leveraging,chen2018harnessing,chi2019nonconvex,zhang2019leverage,zhang2021Cost,pmlr-v151-farias22a,pmlr-v151-tong22a}.
The primary objective for this problem is to reconstruct a low-rank matrix from a limited number of observations. These observations are obtained through a linear map, which consists of the measurement matrices. To be more specific, the measurements for a low-rank matrix 
$\bX \in \C^{M\times N}$ are give by the following
\begin{align}\label{eq:low-rank mode}
  y_k &=\frac{1}{\sqrt{K}}\langle \bA_k, \bX \rangle +z_k, k=1,2,\ldots,K,
\end{align}
where $\bA_k\in \C^{M\times N}$ are the measurement matrices, $\langle \bA_k, \bX \rangle =\tr(\bA_k^H \bX)$, and $z_k$ is the noise or measurement error.
 The $K$ measurement matrices $\{\bA_k\}_{k=1}^K$ collectively define a linear map, denoted as $\mathcal{A}(\cdot):\mathbb{C}^{M \times N} \rightarrow \mathbb{C}^{K \times 1}$,
where each entry of $\cA(\bX)$ is given by
$ [\cA(\bX) ]_k=\frac{1}{\sqrt{K}}\langle \bA_k, \bX \rangle ,  k=1,2,\ldots, K.$ Thus, the measurement model in \cref{eq:low-rank mode} is  shown in a compact form
\begin{align}\label{eq:low-rank compact}
\by = \cA(\bX) + \bz,
\end{align}
where $\by =[y_1,\ldots,y_K]^T$ and $\bz =[z_1,\ldots,z_K]^T$.

The goal of low-rank matrix recovery is to reconstruct $\bX$ from the linear map $\cA(\cdot)$ and measurements $\by$ in \cref{eq:low-rank compact}.
There are various approaches, including both convex and non-convex methods, \citep{recht2010guaranteed,candes2011tight,chen2018harnessing,chi2019nonconvex,ma2018implicit,jain2013low,jain2010guaranteed,zheng2015convergent,tu2016low} can be utilized to fulfill the goal.
The convex methods \citep{recht2010guaranteed,candes2011tight,chen2018harnessing}
utilize the nuclear norm of $\bX$ as a penalty term in their objective functions to promote low-rank solutions. It has been shown that, when the linear map satisfies the restricted isometry property (RIP) with the required RIP constant, these convex methods can guarantee successful recovery in noiseless scenarios or bounded reconstruction errors in the presence of noise.
In addition to convex methods, non-convex techniques \citep{chi2019nonconvex,ma2018implicit,jain2013low,jain2010guaranteed,zheng2015convergent,tu2016low}, such as gradient-based  methods and alternating minimization methods, offer greater computational efficiency .
Notably, these gradient-based  methods, as demonstrated in the works by \cite{chi2019nonconvex,zheng2015convergent,tu2016low},
can ensure convergence with proper initialization when the linear map $\mathcal{A}(\cdot)$ satisfies the RIP with the necessary constant.
Furthermore, some studies \citep{ge2017no,zhang2018much} 
have investigated the presence of spurious local minima in low-rank matrix recovery problems. When there are no spurious local minima, non-convex methods can achieve global minima.
The works by \cite{ge2017no,zhang2018much} have shown that, when the linear map $\cA(\cdot)$ satisfies the RIP with the required constant, the low-rank matrix recovery problem formulated in this manner has no spurious local minima, thereby guaranteeing exact recovery.

As mentioned above, one can find that the RIP of the linear map $\cA(\cdot)$ plays an important role in ensuring the low-rank matrix recovery. In particular, the definition of RIP is presented below.
\begin{Definition}[Standard RIP over Low-Rank Matrices \citep{candes2011tight}]\label{def:standardRIP}
For the set of rank-$r$ matrices, we define the RIP constant
$\delta_r$with respect to operator $\cA(\cdot)$ as the smallest
numbers such that for all $\bX$ of rank at most $r $:
\begin{align*}
  (1-\delta_r)\| \bX \|_F^2 \le \|\cA(\bX)\|_2^2 \le (1+\delta_r)\| \bX \|_F^2.
\end{align*}
\end{Definition}
There are many types of linear map that satisfy the defined RIP above. When the entries of $\bA_k$ are \gls{iid} complex Gaussian entries \citep{recht2010guaranteed,candes2011tight}, or  when the entries of $\bA_k$ are \gls{iid} unit-modulus \citep{zhang2018leveraging}, where each entry is unit-modulus and its phase follows a uniform distribution in the range $[0, 2\pi]$, then the linear map $\cA(\cdot)$
satisfies the RIP with high probability, on the condition that the number of measurements $K\ge c(M+N)r$ for large enough constant $c>0$.

However, in certain scenarios of low-rank matrix recovery, the entries in the measurement matrix are non-\gls{iid} or do not follow the distributions mentioned above, such as the low-rank matrix completion \citep{candes2009exact,candes2010power}, phase retrieval \citep{candes2015phase,ma2018implicit}, and quadratic sensing problem \citep{chen2015exact,cai2015rop}. 
In general, the associated linear map $\mathcal{A(\cdot)}$ in these scenarios does not satisfy the RIP property. To analyze the recovery performance guarantee under these non-RIP scenarios, some variants of RIP are defined, such as incoherence for matrix completion \citep{candes2009exact,candes2010power} and RIP-$\ell_2/\ell_1$ for the rank-one measurements \citep{chen2015exact,cai2015rop}. However, many existing advancements based on RIP, such as the works in \citep{chi2019nonconvex,ma2018implicit,jain2013low,jain2010guaranteed,zheng2015convergent,tu2016low}  , are not applicable for these non-RIP scenarios.

In this paper, our primary focus is on the concept of rank-one measurements as introduced by \cite{cai2015rop}. We aim to demonstrate that when the measurement matrix follows the specified distribution, the associated linear map satisfies the RIP. It is worth noting that the quadratic sensing problem and phase retrieval can be considered special cases of the rank-one projection problem.
For rank-one measurements \citep{cai2015rop,li2019nonconvex}, the measurement matrix $\mathbf{A}_k$ can be represented as an outer product of two vectors,
\begin{align} \label{eq:rank one proj}
\bA_k &= \bu_k \bv_k^H.
\end{align}
However, in general cases where the measurement matrices $\bA_k$ are defined above, it has been established in prior works \citep{cai2015rop,chi2019nonconvex} that the associated linear map does not satisfy the RIP. For example, studies of \cite{chi2019nonconvex,cai2015rop,chen2015exact} have shown that when the entries of both $\bu_k$ and $\bv_k$ are \gls{iid} Gaussian, the associated linear map $\mathcal{A(\cdot)}$ does not satisfy the RIP.
In our work, we impose specific distribution on the random vectors $\bu_k$ and $\bv_k$ instead of  \gls{iid} Gaussian in the existing works \citep{chi2019nonconvex,cai2015rop,chen2015exact}. We show that when the entries in $\mathbf{u}_k$ and $\mathbf{v}_k$ are \gls{iid} unit-modulus, the associated linear map satisfies the RIP with high probability. As far as we know, 
our research marks the first attempt to tackle the challenge of establishing RIP for the linear map in non-\gls{iid} scenarios. Additionally, it lays the foundational framework for proving RIP in various forms of random rank-one measurements.

\section{RIP Analysis of Linear Map}
\subsection{Sufficient Condition of RIP}
For arbitrary $\bu_k$ and $\bv_k$ in \cref{eq:rank one proj}, whether the corresponding linear map $\cA(\cdot)$ satisfies the RIP is challenging to check.
Fortunately, the following theorem provides a sufficient condition on which the linear map $\cA(\cdot)$ satisfies the RIP.

\begin{Theorem}[\cite{candes2011tight}, Theorem 2.3\footnote{It is worth noting that the original result in the work \citep{candes2011tight} is for the real case, i.e., $\cA:\R^{M \times N} \rightarrow \R^{K \times 1}$ and $\bX \in \R^{M \times N}$. However, the result is ready to extend to the complex case.
}] \label{theorem: sufficient}
Let $\cA(\cdot):\C^{M \times N} \rightarrow \C^{K \times 1}$ be a linear map with random measurement matrices obeying the following condition: for any given $\bX \in \C^{M \times N}$ and any fixed $0<\alpha<1$
\begin{align}\label{eq:condition}
\! \! \Pr\left(\left| \| \cA(\bX) \|_2^2 -\| \bX\|_F^2\right| \!\ge \!\alpha\| \bX\|_F^2 \right)\le C \exp(-cK)
\end{align}
holds for fixed constants $C,c>0$ (which may depend on $\alpha$). Then if $K\ge D\max\{M,N \}r$, the linear map $\cA(\cdot)$ satisfies the RIP with constant $\delta_r>0$ with probability exceeding $1-Ce^{-dK}$ for fixed constants $D,d>0$.
\end{Theorem}

Without loss of generality, we assume $\| \bX \|_F = 1$. Therefore, according to \cref{theorem: sufficient}, in order to show the linear map $\cA(\cdot)$ meet RIP, we need to prove the following probability
\begin{align}
\Pr\left (\left|\| \cA(\bX) \|_2^2 - 1\right| \ge \alpha\right) \label{eq:pro bound}
\end{align}
is close to zero. Note that the probability is taken over the linear map $\cA(\cdot)$ and the $\bX$ is fixed and arbitrary.

For the linear map $\cA(\cdot)$  where the entries in $\bA_k$ are drawn from \gls{iid} $\cC\cN(0,1)$,  it is easy to verify that the condition in \cref{eq:condition} holds. Therefore, the corresponding linear map satisfies the RIP with high probability.
This is also consistent with the real case where $\bA_k$ are drawn according to \gls{iid} $\cN(0,1)$ in the following.
\begin{Remark}
[\cite{recht2010guaranteed,candes2011tight}]
If the entries of $\bA_k$ are \gls{iid} Gaussian entries $\cN(0,1)$, then $\cA(\cdot)$ satisfies the $r$-RIP with RIP constant $\delta_r$ with
high probability as $K  \gtrsim (M + N)r/\delta_r^2$.
\end{Remark}

Moreover, when the entries in $\bA_k$ of the linear map are \gls{iid} \citep{recht2010guaranteed,candes2011tight,zhang2018leveraging} , the central limit theorem can be applied to approximately verify the sufficient condition in \cref{eq:condition}. 
However, for the rank-one model in \cref{eq:rank one proj},
the entries of measurement matrix $\bA_k$  are dependent. Due to this dependence, 
the standard RIP in \cref{def:standardRIP} may not hold for the  general $\bu_k$ and $\bv_k$.
For example, when  the entries of $\mathbf{u}_k$ and $\mathbf{v}_k$ are \gls{iid} Gaussian, the RIP does not hold in this scenario because the $\bu^H \bX \bv$  involves fourth moments of Gaussian variable \citep{cai2015rop,kueng2017low,candes2013phaselift}.

To evaluate the concentration property of the linear map for this dependent and rank-one measurement model in \cref{eq:rank one proj},
some alternative conditions, such as RIP-$\ell_1/\ell_1$ \citep{candes2013phaselift} and RIP-$\ell_2/\ell_1$  \citep{chen2015exact,cai2015rop}, have been proposed. These studies demonstrate that the convex methods can ensure the exact recovery based on these alternative conditions.
However, in the context of non-convex analyses, techniques like the alternating minimization method \citep{jain2013low}, singular value projection method \citep{jain2010guaranteed}, and other local optimal analysis \citep{ge2017no,chi2019nonconvex,ma2018implicit}, these variants of RIP \citep{candes2013phaselift,chen2015exact,cai2015rop} are not applicable,  because these analysis are based on the standard RIP. 
Thus, this highlights the crucial importance of standard RIP in \cref{def:standardRIP} compared to its variants.

It is indeed a well-established fact  that the linear map $\cA(\cdot)$ with general setting for $\bu_k$ and $\bv_k$ in \cref{eq:rank one proj} may not satisfy the standard RIP. However, in this paper, we find that
 if we impose some specific design for $\bu_k$ and $\bv_k$,  
 it becomes possible to attain the standard RIP for the designed linear map. The main result of the paper is presented  in the following theorem.
 
 \begin{Theorem} \label{theorem:RIP for unit}
Suppose the measurement matrix $\bA_k=\bu_k \bv_k^H \in \C^{M \times N}$, where $\bu_k \in \C^{M\times 1} ,\bv_k\in \C^{N\times 1} $ are given in the following
\begin{align}
\bu_k &= [e^{j\theta_{k,1}},\ldots,e^{j\theta_{k,M}}]^T,  \nonumber\\
\bv_k &= [e^{j\phi_{k,1}},\ldots,e^{j\phi_{k,N}}]^T, \label{expression uk vk}
\end{align}
with $\theta_{k,m}, \forall m$ and $\phi_{k,n}, \forall n$ being \gls{iid} from a uniform distribution on $[0,2\pi]$. For the linear map $\cA(\cdot):\C^{M\times N} \rightarrow \C^{K \times1}$ generated by 
$\{ \bA_k\}_{k=1}^K$, where 
\begin{align*}
[\cA(\bX) ]_k&=\frac{1}{\sqrt{K}}\langle \bA_k, \bX \rangle ,\\
&=\frac{1}{\sqrt{K}} \bu_k^H \bX \bv_k,
\forall k=1,2,\ldots, K,
\end{align*}
it satisfies RIP with high probability as long as the number of measurements
$K\ge D\max\{M , N\} r $ for some large enough constant $D$.
\end{Theorem}
 
Intuitively, the reason that the standard RIP for the linear map \cref{expression uk vk} holds is because the entries in $\bu_k$ and $\bv_k$ are unit-modulus, which are bounded compared to the scenario where $\bu_k$ and $\bv_k$ are \gls{iid} Gaussian \citep{chen2015exact,cai2015rop}. Moreover, they experience some special symmetric statistical property compared to \gls{iid} Gaussian scenario, which enables us to prove the RIP  in the following sections. Before delving into details of the proof, we first discuss the applications of the measurement model outlined in \cref{expression uk vk}.

Compared to the \gls{iid} measurement matrix $\bA_k$, the rank-one measurements can offer enhanced storage efficiency for the linear map, as demonstrated by \citep{cai2015rop}. Moreover, within the context of rank-one measurements,  the designed unit-modulus setting in \cref{expression uk vk} can further save the storage of  the measurement matrices, as opposed to case of \gls{iid} Gaussian $\bu_k$ and $\bv_k$.
The reason behind this efficiency is that, for the unit-modulus setting described in \cref{expression uk vk}, it is necessary to store only the phases of the vectors $\bu_k$ and $\bv_k$. In contrast, for the \gls{iid} Gaussian $\bu_k$ and $\bv_k$, one must preserve both the magnitudes and phases of these vectors to accurately construct the measurement matrix $\bA_k$.
Most importantly, based on the established results in \cref{theorem:RIP for unit}, the proposed unit-modulus rank-one measurements are applicable for many RIP-based algorithms or analysis \citep{jain2013low,jain2010guaranteed,ge2017no,chi2019nonconvex,ma2018implicit}, making the rank-one unit-modulus measurements a promising option for the matrix recovery task. 
Therefore, building on the advantages highlighted earlier, 
the rank-one measurement model with unit-modulus vectors in \cref{expression uk vk} has widespread applications in the field of low-rank matrix recovery, especially when the configuration of the measurement matrices is applicable, such as  channel estimation in communication systems \citep{el2014spatially,zhang2018leveraging,zhang2020sequential}, phase retrieval \citep{candes2013phaselift,ma2018implicit},  covariance estimation \citep{chen2015exact}, and X-ray crystallography \citep{shechtman2015phase}.

\subsection{Inequalities of Tail Bounds }

To establish the fact that the linear map in \cref{theorem:RIP for unit} satisfies the RIP, we need to evaluate the probability of the event in \cref{eq:pro bound}.
After straightforward manipulations, one can find that $\Eb[\| \by \|_2^2] = 1$.
Thus, the bound in \cref{eq:pro bound} is about the tail bound. When the probability is small, it means that the value of $\| \by\|_2^2$ is strictly concentrated around its expected value.
In the following, we will evaluate the upper and lower tail bounds, respectively.

For the upper tail bound, due to the independence among the entries of $\cA(\bX)$, one can apply Chernoff bound for any $h>0$,
\begin{align*}
\Pr(\| \cA(\bX)\|_2^2\ge (1+\alpha)) 
&= \Pr(h K \| \cA(\bX)\|_2^2\ge h K (1+\alpha)) \nonumber \\
&= \Pr(e^{h K \| \cA(\bX) \|_2^2}\ge e^{h  K(1+\alpha)}) \nonumber \\
&\le  \Eb[e^{h  K\| \cA(\bX) \|_2^2}] e^{-h K (1+\alpha)}.
\end{align*}
Since the second part of the \gls{RHS} of the inequality above, i.e., $e^{-h K (1+\alpha)}$, goes to zero as $K$ goes to infinity, we focus on the first part,
\begin{align*}
\Eb[e^{h  K \| \cA(\bX) \|_2^2)}]&=\Eb[e^{h \sum_{k=1} {| \langle \bA_k , \bX\rangle|^2}}]\\
& = \left(\Eb[e^{h   {|\langle \bA_1 , \bX\rangle|^2}}]\right)^K\\
& = \left( \Eb[e^{{h}{|\bu_1^H  \bX\bv_1|^2}}] \right)^K.
\end{align*}
For convenience, we ignore the subscript  the subscripts of $\bu$ and $\bv$.
Using Taylor series for $e^{h  K {|\langle \bA_1 , \bX\rangle|^2}}$ gives
\begin{align}
\Eb[e^{{h}{|\bu^H  \bX\bv|^2}}]  &=\Eb\left[ \sum_{t=0}^{\infty} \frac{h^t}{t !}|\bu^H \bX\bv|^{2t}\right] \nonumber \\ &= \sum_{t=0}^{\infty} \frac{h^t}{t !} \Eb[ |\bu^H \bX\bv|^{2t}].  \label{eq:chern upper}
\end{align}

For the lower tail bound in \cref{eq:pro bound}, we have
\begin{align*}
\Pr(\| \cA(\bX) \|_2^2\le (1-\alpha)) 
&= \Pr(- h K\| \cA(\bX) \|_2^2\ge  -h K(1-\alpha))  \\
&\le  \Eb[e^{-h K\| \cA(\bX) \|_2^2}] e^{h K(1-\alpha)}\\
&=\left( \Eb[e^{-h{|\bu^H  \bX\bv|^2}}]\right)^K  e^{h K(1-\alpha)} .
\end{align*}
Similarly, using the Taylor series yields
\begin{align}
\Eb[e^{-h{|\bu^H  \bX\bv|^2}}] &= \Eb\left[ \sum_{t=0}^{\infty} \frac{(-h)^t}{t !}|\bu^H \bX\bv|^{2t}\right] \nonumber \\
&= \sum_{t=0}^{\infty} \frac{(-h)^t}{t !} \Eb[ |\bu^H \bX\bv|^{2t}]  . \label{eq:chern lower}
\end{align}

In summary, we have the following upper tail probability bound
\begin{align}
\Pr(\| \cA(\bX) \|_2^2\ge (1+\alpha)) \le 
 \left(\sum_{t=0}^{\infty} \frac{h^t}{t !} \Eb[ |\bu^H \bX\bv|^{2t}]\right)^K  e^{-h K (1+\alpha)}, \label{eq: upper bound}
\end{align}
and the lower tail probability bound
\begin{align} 
\Pr(\| \cA(\bX) \|_2^2\le (1-\alpha))  \le 
\left( \sum_{t=0}^{\infty} \frac{(-h)^t}{t !} \Eb[ |\bu^H \bX\bv|^{2t}]\right)^K  e^{h K(1-\alpha)}. \label{eq: lower bound}
\end{align}

To check whether the sufficient condition in \cref{eq:condition} holds for the linear map $\cA(\cdot)$, we need to evaluate the values in \cref{eq: upper bound} and \cref{eq: lower bound}. By observing the \gls{RHS} of \cref{eq: upper bound} and \cref{eq: lower bound}, one can find that the key is to calculate $ \Eb[ |\bu^H \bX\bv|^{2t}]$.

\section{Connection with the All-One Matrix} \label{sect:All One}
The value of $ \Eb[ |\bu^H \bX\bv|^{2t}]$ depends on the realizations of  $\bX$, which is challenging to manipulate. To handle this, we first focus on a specific $\bX$, where $\bX =\frac{1}{\sqrt{MN}} \bone \bone^T$, then establish a relationship between $ \Eb[ |\bu^H \bX\bv|^{2t}]$ and $ \Eb[ |\bu^H \bone \bone^T\bv|^{2t}]$.
In particular, when $\bX =\frac{1}{\sqrt{MN}} \bone \bone^T$, we have that
\begin{align}
 \Eb\left[ \left|\bu^H \frac{1}{\sqrt{MN}}\bone \bone^T\bv\right|^{2t}\right]
 = 
  \frac{1}{M^tN^t}\Eb\left[\left|\sum_{m=1} \bu_m^*\right|^{2t}\right]\Eb\left[\left|\sum_{n=1} \bv_n \right|^{2t}\right]. \label{eq:bound for one}
 \end{align}
Compared to $ \Eb[ |\bu^H \bX\bv|^{2t}]$,
the value in \cref{eq:bound for one} only depends on the random vector $\bu$ and $\bv$, which is applicable to derive a bound for it.
We first provide some preliminaries, and all their proofs are attached in \cref{app:B} of the supplementary materials.

First of all, the following lemma is about the maximization of summation of combinations.
\begin{Lemma} \label{lem:poly max}
 Suppose $0 \le p \le 1$ and $n\ge 0$,  the summation below
\begin{align*}
  s_n(p) = \sum_{k=0}^{n} \binom{n}{k}^2 p^k (1-p)^{n-k}
\end{align*}
is maximized when $p=1/2$.
\end{Lemma}
Since the summation term $p+(1-p)=1$, the \cref{lem:poly max} shows that the average value, i.e., $p=1/2$, achieves the maximum of $s_n(p)$. The results in \cref{lem:poly max} will be utilized to compare the value of  $ \Eb[ |\bu^H \bX\bv|^{2t}]$ with $ \Eb[ |\bu^H \bone \bone^T\bv|^{2t}]$.
As a straightforward extension, when $p+(T-p)=T$, the following Corollary holds.

\begin{Corollary} \label{lem leb poly}
 Suppose $0 \le p \le T$ and $n\ge 0$,  the summation below
\begin{align*}
  s_n(p)  = \sum_{k=0}^{n} \binom{n}{k}^2 p^k (T-p)^{n-k}
\end{align*}
is maximized when $p=0.5 T$.
\end{Corollary}

Before comparing the values of $ \Eb[ |\bu^H \bX\bv|^{2t}]$ and $ \Eb[ |\bu^H \bone \bone^T\bv|^{2t}]$, 
we  start from the simplified case where $\bx \in \C^{N\times 1}$ and disregard the vector $\bu$. Specifically, we evaluate the values of $\Eb\left[    |  \bx ^T \bv|^{2m} \right]$ and $\Eb\left[    |  \bone ^T \bv|^{2m} \right], \forall m$.
By doing so, we lay the foundation for extending these findings to  a more general settingby incorporating additional considerations

\begin{Theorem} \label{lem: vector case}
Suppose $\bv = [e^{j\phi_{1}},\ldots,e^{j\phi_{N}}]^T$
with $\phi_{n}, \forall n$ being \gls{iid}  from a uniform distribution on $[0,2\pi]$.
 If $\bx \in \C^{N \times 1}$ with $\| \bx\|_2=1$, then for any $m\ge 0$, $\Eb\left[    |  \bx ^T \bv|^{2m} \right]$ is maximized when
 $x_n = \frac{1}{\sqrt{N}}, \forall n$.
\end{Theorem}
Thus, according to \cref{lem: vector case}, for any $\| \bx\|_2=1$, one has the following inequality,
\begin{align*}
\Eb[    |  \bx ^T \bv|^{2m} ] \le \frac{1}{N^m}\Eb[    |  \bone ^T \bv|^{2m} ], \forall m.
\end{align*}
Furthermore, in order to extend the vector-case results in \cref{lem: vector case} to a more general setting, we need the following \cref{lem: vector case c,lem:holder N2,lem:holder N general} as preliminaries.
\begin{Lemma} \label{lem: vector case c}
Suppose  $t>0$  and $k_m\ge 0$ are integers, and  non-negative $c_m\in \R, \forall m=1,2,\ldots,M$, with $\sum_{m=1}^{M} c_m^2 =1$, then the following holds
\begin{align} \label{eq:lem vector case c}
\sum_{k_1+\ldots+k_M=t} {\binom{t}{k_1, \ldots,k_M}}^2    \prod_{m=1}^{M}c_m^{2k_m} \le 
\sum_{k_1+\ldots+k_M=t}  M^{-t} {\binom{t}{k_1, \ldots,k_M}}^2,
\end{align}
where the equality holds when $c_m=1/\sqrt{M}$.
\end{Lemma}
The results in \cref{lem: vector case c} mean that the summation about the combinatorial expression is maximized when the values of $c_m,\forall m,$ are equivalent, which is the extension of result in the one-variable case of \cref{lem leb poly}.

\begin{Lemma} \label{lem:holder N2}
Suppose non-negative random variables $X_1,X_2$, for any $k_1,k_2\ge 0$ and $1/p+1/q=1$ with $p,q \in [1, +\infty)$, then we have
\begin{align} \label{eq:holder N2 pq}
\Eb \left[ X_1^{k_1} X_2^{k_2}\right] &\le (\Eb[X_1^{k_1 p}])^{1/p} (\Eb[X_2^{k_2 q}])^{1/q}.
\end{align}
In particular,
\begin{align} \label{eq:holder N2 max}
\Eb [ X_1^{k_1} X_2^{k_2}] & \le \max  \{  \Eb [ X_1^{k_1+k_2} ], \Eb [ X_2^{k_1+k_2} ] \}.
\end{align}

\end{Lemma}

The results in \cref{lem:holder N2} are to bound $\Eb [ X_1^{k_1} X_2^{k_2}]$ by the product of expectations, where the latter is more applicable to handle.
Then, the following lemma is an extension of the results in \cref{lem:holder N2}, where there are $N$ random variables.

\begin{Lemma} \label{lem:holder N general}
Suppose non-negative random variables $X_1,X_2, \ldots, X_N$, for any $k_n\ge 0$, then the following inequality about expectation holds
\begin{align}
\Eb \left[\prod_{n=1}^{N} X_n^{k_n}\right] &\le \prod_{n=1}^N\left(\Eb\left[X_{n}^{t}\right]\right)^{k_n/t} \nonumber\\
& \le \max_n \Eb\left[X_n^t\right] ,\label{eq:holder N general max}
\end{align}
where $t=\sum_{n=1}^{N} k_n$.

\end{Lemma}

With the results in \cref{lem: vector case c,lem:holder N2,lem:holder N general}, we are now ready to compare the values of $ \Eb[ |\bu^H \bX\bv|^{2t}]$ and $ \frac{1}{M^tN^t} \Eb[ |\bu^H \bone \bone^T\bv|^{2t}]$ in the following theorem, which is proved in \cref{app:A} of the supplementary materials.
\begin{Theorem} \label{lemma:all one ineq}
For any matrix $\bX\in \C^{M\times N}$ with $\|\bX\|_F = 1$, the following
\begin{align*}
  \Eb[ |\bu^H \bX \bv|^{2t}]  \le  \frac{1}{M^tN^t} \Eb[ |\bu^H \bone \bone^T\bv|^{2t}]
\end{align*}
 holds for any non-negative integer $t$.
 Here, $\bu\in\C^{M\times 1}$ and $\bv\in\C^{N\times 1}$ are random vectors given by
 \begin{align*}
\bu = [e^{j\theta_{1}},\ldots,e^{j\theta_{M}}]^T, ~\bv = [e^{j\phi_{1}},\ldots,e^{j\phi_{N}}]^T
\end{align*}
with $\theta_{m},\forall m$ and $\phi_{n} ,\forall n$ being i.i.d in $[0,2\pi]$ .
\end{Theorem}

Thus, the results in \cref{lemma:all one ineq} establish a relationship between $ \Eb[ |\bu^H \bX\bv|^{2t}]$ and $ \Eb[ |\bu^H \bone \bone^T\bv|^{2t}]$, where the latter the only depends on the random vector $\bu$ and $\bv$.
Therefore, to further proceed with \cref{lemma:all one ineq} and obtain a valid bound for  $\Eb[ |\bu^H \bX \bv|^{2t}] $ in \cref{eq: upper bound} and \cref{eq: lower bound}, 
it necessary to assess the value of $ \Eb[ |\bu^H \bone \bone^T\bv|^{2t}]$. 
Due to the independence between $\bu$ and $\bv$,  one can express this as $ \Eb[ |\bu^H \bone \bone^T\bv|^{2t}] =  \Eb[ |\bu^H \bone|^{2t} ] \Eb[ \bone^T\bv|^{2t}]$. In this context, the following proposition provides a valuable bound for both $\Eb[ |\bu^H \bone|^{2t} ] $ and  $\Eb[ \bone^T\bv|^{2t}]$.

\begin{Proposition}\label{pro:Abiean suqres}
Since the entries in $\bu$ and $\bv$ are \gls{iid}, one can check that
\begin{align*}
\Eb\left[\left|\bu^H \bone\right|^{2t}\right] 
&= 
\Eb\left[ \left( u_1+,\ldots,+u_M\right)^t \left( u_1^*+,\ldots,+u_M^*\right)^t \right]\\
& = \sum_{t_1+\ldots+t_M=t} {\binom{t}{t_1, \ldots,t_M}}^2.    
\end{align*}
One can find that the value above is the number of abelian squares of length $2t$ over an alphabet  with $M$ letters  \citep{richmond2008counting}, denoted as $g(t,M)$. Similarly, the value of $\Eb\left[ |\bone^T\bv|^{2t}\right]$ is given by
\begin{align*}
\Eb\left[ |\bone^T\bv|^{2t}\right]
 &= \sum_{t_1+\ldots+t_N=t} {\binom{t}{t_1, \ldots,t_N}}^2 \\
   &=g(t,N).
\end{align*}
Now, the bounds of $g(t,N)$ and $g(t,M)$ are of interest. 
Based on the results by \cite{richmond2008counting},  the values of $g(t,M)$  and $g(t,N)$ are bounded by
\begin{align*}
g(t,M) &\le C_1 M^{2t}t^{(1-M)/2}, \\
g(t,N) &\le C_2 N^{2t}t^{(1-N)/2},
\end{align*}
where $C_1$ and $C_2$ are two constants. Therefore,  the value of $ \Eb[ |\bu^H \bone \bone^T\bv/{\sqrt{MN}}|^{2t}]$ is upper bounded by
\begin{align} \label{bound temp 1}
\Eb\left[\left|\bu^H \frac{\bone \bone^T }{\sqrt{MN}} \bv\right|^{2t}\right] \le C M^t N^t t^{1-\frac{M+N}{2}},
\end{align}
where $C=C_1 C_2$ is a constant.
\end{Proposition}

\section{RIP of Rank-One Unit-Modulus Measurements}
In this section, we prove the RIP of rank-one measurements with unit-modulus vectors in \cref{theorem:RIP for unit}. Recall that it is essential to evaluate the bounds in \cref{eq: upper bound} and \cref{eq: lower bound} to show  they satisfy the sufficient condition in \cref{theorem: sufficient}.
Based on the established results in \cref{sect:All One}, we can simplify the upper and lower tail bounds in \cref{eq: upper bound} and \cref{eq: lower bound}, respectively, as shown in the following theorem.

\begin{Theorem} \label{theorem:tail  c}
For the linear map $\cA(\cdot)$ defined in \cref{theorem:RIP for unit}, we have the following upper tail probability bound for any $\bX \in \C^{M\times N}$ with $\| \bX \|_F = 1$,
\begin{align} \label{eq:upper final}
\Pr(\| \cA(\bX) \|_2^2\ge (1+\alpha))  \le  e^{-c_1K},
\end{align}
where $c_1>0$ is a constant depending on $\alpha$.
In addition, the lower tail probability bound is given by
\begin{align}\label{eq:lower final}
\Pr(\| \cA(\bX) \|_2^2\le (1-\alpha))  \le   e^{-c_2 K},
\end{align}
where $c_2>0$ is  also a constant depending on $\alpha$.
\end{Theorem}
According to  the results in \cref{theorem:tail  c}, 
it is evident that both the upper and lower tail bounds exhibit an exponential decrease as the number of measurements $K$. This observation leads us to verify the sufficient condition for the RIP outlined in \cref{theorem: sufficient}. Consequently, the linear map associated with random unit-modulus vectors satisfies the RIP with high probability, as shown in \cref{theorem:RIP for unit}. In the following, we provide a comprehensive proof of \cref{theorem:tail  c}.

\textit{Proof of \cref{theorem:tail  c}.}
To prove the upper tail bound in \cref{eq:upper final}, we combine the results in \cref{eq: upper bound} and \cref{lemma:all one ineq},
\begin{align*}
\Pr(\| \cA(\bX)  \|_2^2\ge (1+\alpha))
 &\le  \left( \sum_{t=0}^{\infty} \frac{h^t}{t !} \Eb[ |\bu^H \bX\bv|^{2t}] \right)^Ke^{-h K(1+\alpha)} \\
& \le\left( \sum_{t=0}^{\infty} \frac{h^t}{t !} \Eb\left[\left|\bu^H \frac{\bone \bone^T }{\sqrt{MN}} \bv\right|^{2t}\right] \right)^Ke^{-h K(1+\alpha)}.
\end{align*}
Then from \cref{pro:Abiean suqres}, we have
\begin{align}
&\Pr(\| \cA(\bX)  \|_2^2\ge (1+\alpha)) \nonumber \\
&{\le} \Bigg( 1+    h+2h^2-\frac{2M+2N-1}{2MN}h^2+ 
  \sum_{t=3}^{\infty} \frac{h^t}{t !}  C M^t N^t t^{1-\frac{M+N}{2}} \Bigg)^Ke^{-h K(1+\alpha)}. \label{eq:tail upper 1}
\end{align}
In the following, we need to choose a $h$ which makes the bound above tight.
Note that there exists $Z>0$, which makes the following hold  for any $0<h \le Z \alpha$,
\begin{align}
1+    h+2h^2-\frac{2M+2N-1}{2MN}h^2+ 
\sum_{t=3}^{\infty} \frac{h^t}{t !}  C M^t N^t t^{1-\frac{M+N}{2}} \le  1+    h+2h^2. \label{eq:tail approx 1}
\end{align}

For convenience, we define
\begin{align*}
f(h) = (1+h+2h^2)e^{-h(1+\alpha)}.
\end{align*}
Thus, according to \cref{eq:tail upper 1} and \cref{eq:tail approx 1}, for any $0<h \le Z \alpha$, the tail probability in \cref{eq:tail upper 1} is bounded as follows
\begin{align} \label{eq:upper tail 2}
\Pr(\| \cA(\bX)  \|_2^2\ge (1+\alpha)) \le (f(h))^K.
\end{align}
Comparing \cref{eq:upper tail 2} with the sufficient condition in \cref{eq:condition}, we need to prove that the above expression has the form of $C\exp(-cK)$.
 In other words, we need to show there exists a $h_1\in (0,Z\alpha]$ such that $
 f(h_1)  < 1$, then  $(f(h_1))^K$ converges to zero as $K$ goes to infinity.

Note that $f(0)=1$. Thus, in order to show there exists $h_1$ such that $f(h_1)<1$, it is sufficient to show the first derivative of $f$ at $0$ is negative, i.e., $f'(0)<0$, which is obviously true. Therefore, we choose $h=h_1$, the expression in \cref{eq:upper tail 2} is rewritten as
\begin{align*}
\Pr(\| \cA(\bX)  \|_2^2\ge (1+\alpha)) &\le ( 1+    h_1+2 h_1^2)^K e^{- h_1 K(1+\alpha)}\\
&= e^{-c_1K},
\end{align*}
where the constant $c_1=-\ln(f(h_1))=-\ln((1+h_1+2h_1^2)e^{-h_1(1+\alpha)}) >0$. Thus, the bound for the upper tail probability \cref{eq:upper final} is proved.

For the lower tail probability in \cref{eq:lower final},  we have
\begin{align}
\Pr(\| \cA(\bX)  \|_2^2\le (1-\alpha))  
&\le \sum_{t=0}^{\infty} \frac{(-h)^t}{t !} \Eb[ |\bu^H \bX\bv|^{2t}]   e^{h K(1-\alpha)} \nonumber \\
& \le \left(  1-h + 2h^2\Eb[ |\bu^H  \bX\bv|^4] \right) ^K  e^{h K(1-\alpha)} \nonumber \\
 &\le \left(  1-h + 2h^2\Eb\left[\left|\bu^H \frac{\bone \bone^T }{\sqrt{MN}}\bv\right|^4\right] \right) ^K  e^{h K(1-\alpha)} \nonumber \\
&\le  ( 1-    h+2h^2)^K e^{ hK(1-\alpha)}. \label{eq:lower tail 1}
\end{align}
Similarly, we need to prove that the above expression has the form of $C\exp(-cK)$ in \cref{theorem: sufficient}, 
where $C$ and $c$ are constants. The straightforward method is to minimize the value above with respect to $h$. Here, for simplicity, we just let $h=\frac{1}{4}$. Then, the lower tail probability in \cref{eq:lower tail 1} is  bounded by
\begin{align*}
\Pr(\| \cA(\bX)  \|_2^2\le (1-\alpha)) \le 0.875^{K} e^{\frac{1}{4}K(1-\alpha)}
\le e^{-c_2K},
\end{align*}
where  $c_2=1.83+({1}/{4})\alpha >0$. 
Therefore, the bound for the lower tail probability \cref{eq:lower final} is proved.
\qed

Based on the results of \cref{theorem:tail  c}, we finally provide the proof of the \cref{theorem:RIP for unit}, and show that the linear map associated with the random unit-modulus vectors satisfies the RIP with high probability.

\textit{Proof of \cref{theorem:RIP for unit}.}
By utilizing the union bound for the probability in \cref{eq:pro bound}, we have
\begin{align*}
&\Pr\left (\left|\|\cA(\bX)\|_2^2 - 1\right| \ge \alpha\right) \\&= \Pr\left (\| \cA(\bX)\|_2^2 \ge 1+ \alpha\right)+\Pr\left (\| \cA(\bX)\|_2^2 \le 1- \alpha\right).
\end{align*}
According to the results in\cref{theorem:tail c}, combining the upper and tail bounds together  gives
\begin{align*}
\Pr(\left| \| \cA(\bX) \|_2^2- 1\right| \le \alpha ) &\le   e^{-c_1K} +  e^{-c_2K}\\
&\le 2 e^{-cK},
\end{align*}
where $c =\min(c_1,c_2)$. Furthermore, according to \cref{theorem: sufficient},   if the number of measurements $K\ge D\max\{M,N \}r$, the linear map $\cA(\cdot)$ satisfies the RIP with isometry constant $\delta_r >0$ with probability exceeding $1-2e^{-dK}$ for fixed constants $D,d>0$.
\qed

\begin{figure*}[!htbp]
\centering  
\begin{subfigure}[b]{0.37\textwidth}
\centering    
\includegraphics[width=0.9\textwidth]{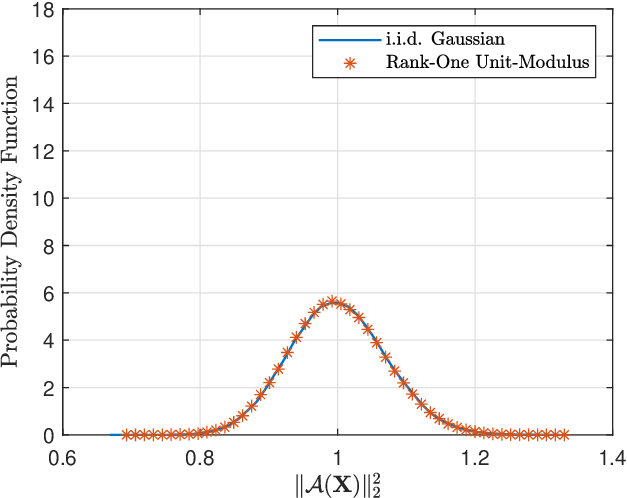}
\caption{ $K=200$}
\end{subfigure}
\begin{subfigure}[b]{0.37\textwidth}
\centering    
\includegraphics[width=0.9\textwidth]{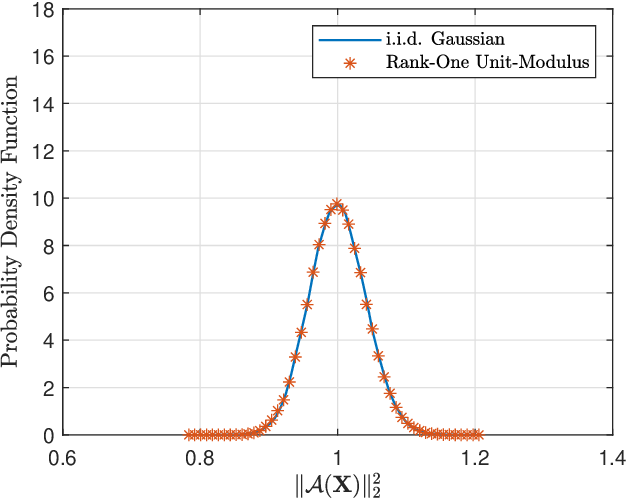}
\caption{ $K=600$}
\end{subfigure}
\par
\begin{subfigure}[b]{0.37\textwidth}
\centering    
\includegraphics[width=0.9\textwidth]{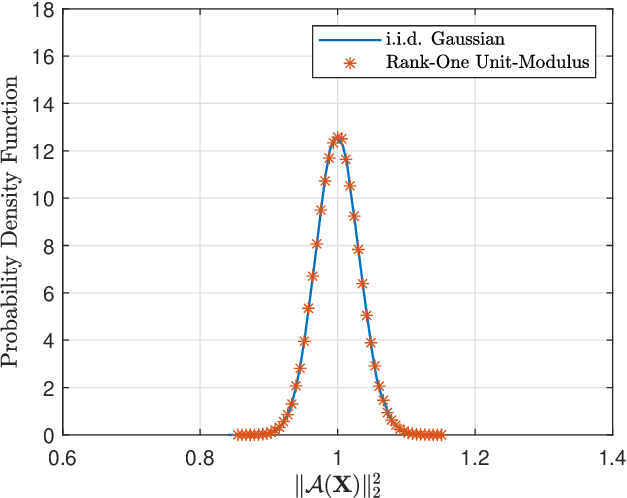}
\caption{ $K=1000$}
\end{subfigure}
\begin{subfigure}[b]{0.37\textwidth}
\centering    
\includegraphics[width=0.9\textwidth]{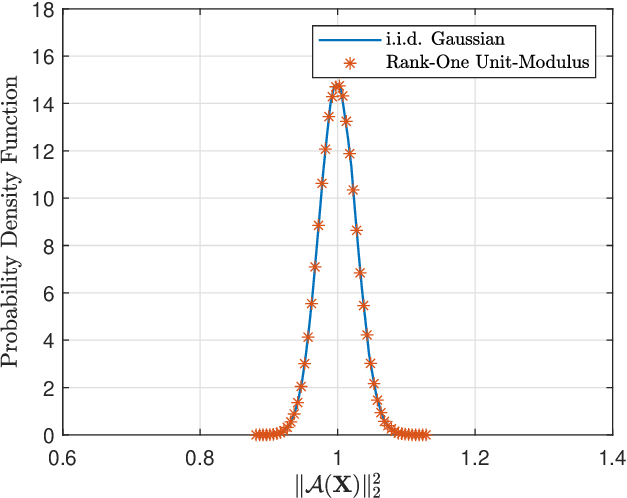}
\caption{ $K=1400$}
\end{subfigure}
\caption{Probability Density Function of $\| \cA(\bX)\|_2^2$ with Different Number of Measurements}  
\label{fig:different K}    
\end{figure*}

\section{Numerical Experiments}

In this section, we verify that the linear map generated by the random unit-modulus satisfies the RIP. Subsequently, we assess the recovery performance of low-rank matrices by employing this linear map.

Directly validating the RIP of the linear map is known to be NP-hard.
Hence, we opt to demonstrate that the sufficient condition for RIP as outlined in \cref{theorem: sufficient} holds true. This condition suggests that the value of $\| \mathcal{A}(\mathbf{X})\|_F^2$ associated with unit-modulus vectors is highly concentrated around its expected value, which is $\|\mathbf{X} \|_F^2$.
To illustrate this, we conduct an experiment as presented in \cref{fig:different K}. In this experiment, we randomly generate a fixed $\mathbf{X} \in \mathbb{C}^{40\times 80}$ with $\|\mathbf{X} \|_F = 1$ and examine the probability density functions of $\| \mathcal{A}(\mathbf{X})\|_F^2$ by using two types of linear map. The first is generated by the rank-one model using random unit-modulus vectors, while the second serves as the benchmark and is based on \gls{iid} Gaussian $\mathbf{A}_k$.


As depicted in \cref{fig:different K}, it is evident that the value of $\| \cA(\bX)\|_F^2$ generated by rank-one measurements with random unit-modulus vectors is concentrated around  its expectation $\| \bX \|_F^2$ for different number of measurements. Comparing this outcome to the scenario where the entries of measurement matrix $\bA_k$ are \gls{iid} Gaussian, we observe that the probability density function gap between these two scenarios is notably narrow. Therefore, according to \cref{theorem: sufficient}, the linear map employing the random rank-one unit-modulus measurements satisfies the RIP with high probability, similar to the scenario using  \gls{iid} Gaussian $\bA_k$.
Furthermore, as the number of measurements $K$ increases, both the Gaussian and rank-one unit-modulus curves become increasingly tightly concentrated around the expected value $\|\bX \|_F^2$. This is  consistent with the results in \cref{theorem:tail  c}, where the tail bound exponentially decreases with the number of measurements $K$.

\begin{figure}[!t]
\centering  
\includegraphics[width=0.55 \textwidth]{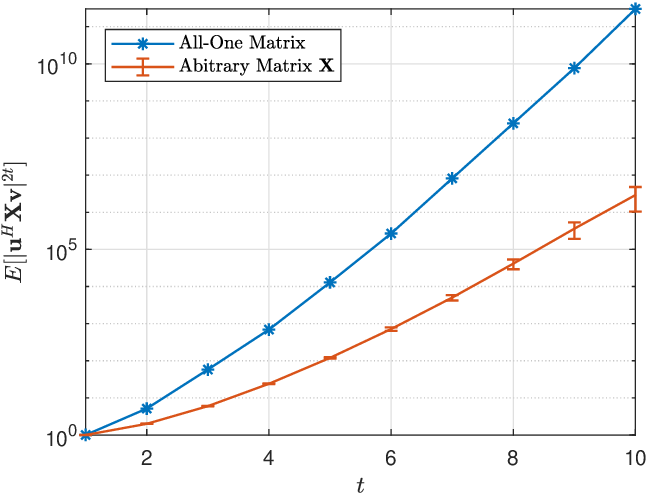}
\caption{Comparison of Two Expectations: All-One Matrix Scenario $\frac{1}{M^tN^t} \Eb[ |\bu^H \bone \bone^T\bv|^{2t}]$ and Arbitrary Matrix Scenario $ \Eb[ |\bu^H \bX\bv|^{2t}]$ } 
\label{fig:all one}   
\end{figure}

Since the results in \cref{lemma:all one ineq} are essential in the derivation of RIP analysis, we   conduct an experiment in \cref{fig:all one} to confirm the correctness of  \cref{lemma:all one ineq}.
 Specifically, in \cref{lemma:all one ineq}, we have established the fact: under the constraint $\|\bX\|_F=1$, the random variable $|\mathbf{u}^H \mathbf{X} \mathbf{v}|$ achieves the largest $(2t)^{th}$ moment when $\mathbf{X} = \frac{1}{\sqrt{MN}}\mathbf{1} \mathbf{1}^T$. In other words,  we conclude that $ \mathbb{E}[ |\mathbf{u}^H \mathbf{X} \mathbf{v}|^{2t}] \le \frac{1}{M^tN^t} \mathbb{E}[ |\mathbf{u}^H \mathbf{1} \mathbf{1}^T\mathbf{v}|^{2t}], \forall t$.
 For the experiment in \cref{fig:all one}, we randomly generate $100$ matrices $\mathbf{X} \in \mathbb{C}^{40 \times 80}$ and empirically calculate $\mathbb{E}[ |\mathbf{u}^H \mathbf{X} \mathbf{v}|^{2t}]$ for each $\bX$ and $t=1,2,\ldots, 10$. The blue line represents the scenario of the all-one matrix, $ \frac{1}{M^tN^t} \mathbb{E}[ |\mathbf{u}^H \mathbf{1} \mathbf{1}^T\mathbf{v}|^{2t}]$. The red curve illustrates the mean and standard deviation for the $100$ realizations of arbitrary $\mathbf{X}$.
 Upon reviewing \cref{fig:all one}, we can readily observe that $\mathbb{E}[ |\mathbf{u}^H \mathbf{X} \mathbf{v}|^{2t}] \le {1}/{(M^tN^t)} \mathbb{E}[ |\mathbf{u}^H \mathbf{1} \mathbf{1}^T\mathbf{v}|^{2t}], \forall t$. Therefore, the value of $\mathbb{E}[ |\mathbf{u}^H \mathbf{X} \mathbf{v}|^{2t}]$ is maximized when the matrix $\mathbf{X}$ has the form of the all-one matrix, i.e., $\mathbf{X} = \frac{1}{\sqrt{MN}}\mathbf{1} \mathbf{1}^T$. In conclusion, these experimental results are in perfect alignment with our analysis in \cref{lemma:all one ineq}.

\begin{figure}[t]
\centering  
\includegraphics[width=0.55\textwidth]{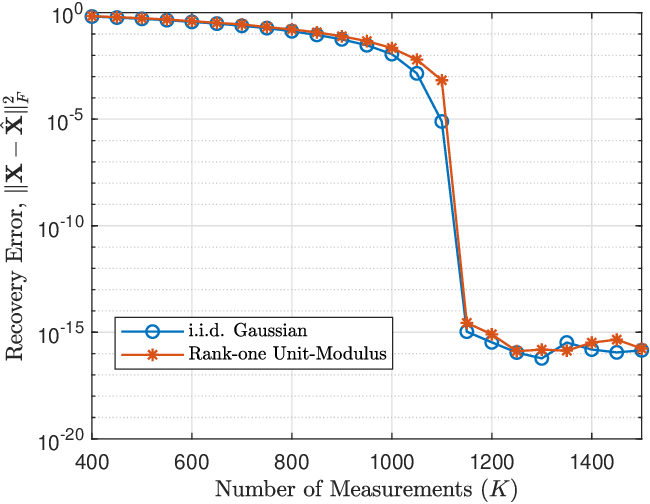}
\caption{Recovery Error of Low-Rank Matrix by Using  Rank-One Unit-Modulus and i.i.d. Gaussian Measurements with Different Number of Measurements} 
\label{fig:recover different K}   
\end{figure}

In \cref{fig:recover different K}, we evaluate the low-rank matrix recovery performance by using the linear map of the rank-one unit-modulus measurements and \gls{iid} measurement matrix $\bA_k$. We randomly generate the target complex low-rank matrix $\bX$ with dimension of $40\times 80$ of rank $r=5$. 
We let the number of measurements $K$ vary from $400$ to $1500$. For fair comparison, we utilize the well-known nuclear norm minimization \citep{recht2010guaranteed} to recover the low-rank matrix from the measurements. The optimization problem is given by
\begin{align*} 
\begin{split}
&\min_{\bX} \| \bX  \|_*   \\
&\text{subject to } \cA(\bX) = \by.
\end{split}
\end{align*}

We can refer to the findings in \citep{recht2010guaranteed} that the recovery error by using \gls{iid} $\bA_k$ decreases with the number of measurements $K$.  
As evident in \cref{fig:recover different K},
the recovery error of the designed linear map exhibits a similar trend as the case of \gls{iid} Gaussian. 
Moreover, observe that there is a sharp transition to near zero error at around $1100$ measurements for these two scenarios, which is consistent with the results in \citep{recht2010guaranteed}.
Overall, the observations in  \cref{fig:recover different K} suggest that  rank-one measurements with unit-modulus vectors achieves a recovery performance similar to that of  \gls{iid} Gaussian $\mathbf{A}_k$. Additional numiercal experiments about the recovery performance by using non-convex matrix recovery algorithms are attached in \cref{app:C} of the supplementary materials.

Furthermore, the use of unit-modulus vectors allows for more efficient memory storage for the linear map, reducing the hardware costs associated with the linear map. As we have analyzed, this type of linear map not only satisfies the RIP but also exhibits similar recovery performance as the case of \gls{iid} measurement matrices, as demonstrated in our experiments. 
Therefore, these advantages position our designed rank-one measurements with unit-modulus vectors as a promising linear map in low-rank matrix recovery.

\section{Conclusion}

In this paper, we have conducted a comprehensive RIP analysis for rank-one measurements with random unit-modulus vectors. The symmetric statistical properties of unit-modulus vectors have allowed us to derive the tail bound which exponentially decrease with the number of measurements . In comparison to the scenario of \gls{iid} measurement matrices, the linear map generated by unit-modulus vectors not only satisfies the RIP but also offers the high memory efficiency. 
These advantageous properties show the potential of rank-one unit-modulus measurements as a highly effective linear map in the field of low-rank matrix recovery.

\bibliography{references}

\onecolumn

\setcounter{section}{0}
\renewcommand{\thesection}{\Alph{section}}

{The supplementary materials contain detailed proofs of the results that are missing in the main paper, and additional experiments are provided as well.}

\section{Proof of  \cref{lemma:all one ineq}} \label{app:A}

After standard manipulations, one can find that
\begin{align*}
  \Eb[ |\bu^H \bX \bv|^{2t}]  \le  \Eb[ |\bu^H \bar{\bX} \bv|^{2t}],
\end{align*}
where each entry in $\bar{\bX} $ is equal to the absolute value of the corresponding entry in $\bX$. Therefore, it remains to show that
\begin{align*}
 \Eb[ |\bu^H \bar{\bX} \bv|^{2t}] \le   \frac{1}{M^tN^t} \Eb[ |\bu^H \bone \bone^T\bv|^{2t}].
\end{align*}
For the easy notation, we let $\bX = \bar{\bX}$.
Denoting $\bX_{m,:}$ as the $m$th row of $\bX$, we have
\begin{align*}
\Eb\left[    |  \bu^H \bX \bv|^{2t} \right]
= \Eb\left[\left(\sum_{m} u_m^* \bX_{m,:}  \bv \right) ^t \left(\sum_{m} u_m \bX_{m,:}  \bv^* \right) ^t \right].
\end{align*}
Then,  according to the multinomial theorem, we have
\begin{align}
&\Eb\left[    |  \bu^H \bX \bv|^{2t} \right] \nonumber \\
&=\Eb\left[ \sum_{k_1+\ldots+k_M=t} \binom{t}{k_1, \ldots,k_M}\prod (u_m^* \bX_{1,:} \bv)^{k_m}     \sum_{k_1'+\ldots+k_M'=t} \binom{t}{k'_1, \ldots,k_M'}\prod (u_m \bX_{m,:}  \bv^*)^{k'_m}  \right] \nonumber \\
&=\Eb\left[ \sum_{k_1+\ldots+k_M=t} {\binom{t}{k_1, \ldots,k_M}}^2\prod | \bX_{m,:}  \bv|^{2k_m}  \right]. \label{supp:eq b1}
\end{align}

For the concise proof, we express $\bX_{m,:}=c_m\bx_m^T $ where $\|\bx_m\|_2 = 1$. Without loss generality, we assume $c_m\ge0, \forall m$. Then we rewrite the expression \cref{supp:eq b1} above as
\begin{align*}
&\Eb\left[ \sum_{k_1+\ldots+k_M=t} {\binom{t}{k_1, \ldots,k_M}}^2\prod | \bX_{m,:}  \bv|^{2k_m}  \right]\\
&=
\Eb\left[ \sum_{k_1+\ldots+k_M=t} {\binom{t}{k_1, \ldots,k_M}}^2\prod | \bx_m^T \bv|^{2k_m} \prod_{i=1}^{M}c_m^{2k_m} \right].
\end{align*}
Then, according to \cref{lem:holder N general}, we have
\begin{align*}
\Eb\left[ \prod | \bx_m^T \bv|^{2k_m} \right] \le \max_{m}\Eb\left[  |\bx_m^T \bv|^{2t} \right].
\end{align*}
Thus, the value of $\Eb\left[    |  \bu^H \bX \bv|^{2t} \right]$ in \cref{supp:eq b1} is upper bounded by
\begin{align*}
\Eb\left[    |  \bu^H \bX \bv|^{2t} \right] &\le
 \sum_{k_1+\ldots+k_M=t} {\binom{t}{k_1, \ldots,k_M}}^2    \max_{m}\Eb\left[  |\bx_m^T \bv|^{2t} \right] \prod_{i=1}^{M}c_m^{2k_m} \\
  &=\Eb\left[    |  \bu^H \bX^* \bv|^{2t} \right],
\end{align*}
where
$\bX^*=[c_1 \bx_{m^*} , c_2 \bx_{m^*} , \cdots , c_M \bx_{m^*}]^T$
and $ m^*=\argmax_m \Eb\left[ | \bX_{m,:}  \bv|^{2t}\right]$.
In other words,
\begin{align}
\Eb\left[    |  \bu^H \bX \bv|^{2t} \right] &\le \Eb\left[    |  \bu^H \bX^* \bv|^{2t} \right]  \nonumber\\
&=  \sum_{k_1+\ldots+k_M=t} {\binom{t}{k_1, \ldots,k_M}}^2    \Eb\left[  |\bx_{m^*}^T \bv|^{2t} \right] \prod_{i=1}^{M}c_m^{2k_m} \nonumber\\
&=   \underbrace{\Eb\left[  |\bx_{m^*}^T \bv|^{2t} \right]}_{\text{first part}} \underbrace{\sum_{k_1+\ldots+k_M=t} {\binom{t}{k_1, \ldots,k_M}}^2    \prod_{i=1}^{M}c_m^{2k_m}}_{\text{second part}}. \label{eq:A proof 1}
\end{align}
By \cref{lem: vector case}, the first part of value in \cref{eq:A proof 1} is maximized when the entries in $\bx_{m^*}$ are equivalent. 
According to \cref{lem: vector case c}, the second part is maximized when $c_m = \sqrt{\frac{1}{M}}$.
Recall that $\bX^*=[c_1 \bx_{m^*} , c_2 \bx_{m^*} , \cdots , c_M \bx_{m^*}]^T$, after substituting the setting of $\bx_{m^*}$ and $c_m$ into the expression of $\bX^*$, one can easily check the corresponding matrix is expressed as $\bX^*=\frac{1}{\sqrt{MN}} \bone \bone^T$. Thus,
we have
\begin{align*}
\Eb\left[    |  \bu^H \bX \bv|^{2t} \right]  \le  \frac{1}{M^tN^t} \Eb[ |\bu^H \bone \bone^T\bv|^{2t}].
\end{align*}
This concludes the proof. \qed

\section{Proof of Preliminaries}\label{app:B}
\subsection{Proof of \cref{lem:poly max}} \label{sect:proof poly max}
Note that the Legendre polynomials $P_n(x)$ is expressed as
\begin{align*}
P_n(x) = \frac{1}{2^n} \sum_{k=0}^n \binom{n}{k}^2 (x-1)^{n-k} (x+1)^k.
\end{align*}
After simple manipulations, the expression of $s_n(p)$ is expressed as
\begin{align*}
s_n(p) = (2p-1)^n P_n\left(\frac{1}{2p-1}\right).
\end{align*}
After performing first-order derivative, one can check that the value of $s_n$ is maximized when $p=1/2$.  \qed

\subsection{Proof of \cref{lem: vector case}} \label{se: ap B2}
 Without loss of generality, we can assume that entries in $\bx$ are non-negative, i.e., $x_n \ge 0, \forall n$. This is because $\Eb\left[    |  \bx ^T \bv|^{2 m} \right] \le \Eb\left[    |  \bar{\bx} ^T \bv|^{2 m} \right]$.
Then, the expression of $\Eb\left[    |  \bx ^T \bv|^{2_m} \right]$ is
\begin{align}
\Eb\left[    |  \bx ^T \bv|^{2m} \right]
 = \Eb\left[\left(x_1v_1+\ldots+  x_N v_N\right) ^m \left(x_1 v_1^*+\ldots + x_N v_N^*\right)^m \right]. \label{eq: A proof 2}
\end{align}
Without loss of generality, we assume $x_1 \neq x_2$. To complete the proof, we will show that by letting $x_1 =x_2= \sqrt{\frac{x_1^2+x_2^2}{2}}$, the expectation in \cref{eq: A proof 2} will increase.
Here, we denote $y = \sum_{n=3}^{N} x_n v_n$, and $t = \sum_{n=3}^{N} x_n^2$. Therefore, we have $x_1^2+x_2^2 = 1-t$.
Then, according to the multinomial theorem, we have
\begin{align*}
\Eb\left[    |  \bx ^T \bv|^{2_m} \right]&= \Eb\left[\left(x_1v_1+x_2v_2+  y \right) ^m \left(x_1 v_1^*+x_2v_2^* + y^*\right)^m \right]\\
& = \Eb \Biggl[\sum_{k_1+ k_2+ k_3=m} \binom{m}{k_1, k_2, k_3}(x_1v_1)^{k_1} (x_2v_2)^{k_2} y^{k_3}\\
&~~~~~~~~\sum_{k_1'+k_2'+ k_3'=m} \binom{m}{k_1', k_2', k_3'}(x_1v_1^*)^{k_1'} (x_2v_2^*)^{k_2'} y^{k_3'}\Biggr].
\end{align*}
Because of the property of expectation, the expression above can be written in
\begin{align*}
 &\sum_{k_1+ k_2+ k_3=m} {\binom{m}{k_1, k_2, k_3}}^2(x_1)^{2k_1} (x_2)^{2k_2} \Eb \left [|y|^{2 k_3}\right]\\
 &=\sum_{k_3=0}^{m} \sum_{k_1+ k_2=m-k_3}{\binom{m}{k_1, k_2, k_3}}^2(x_1)^{2k_1} (x_2)^{2k_2} \Eb \left [|y|^{2 k_3}\right]\\
  &=\sum_{k_3=0}^{m} \sum_{k_1+ k_2=m-k_3}{\binom{m}{k_3, m-k_3}}^2 {\binom{m-k_3}{k_1, k_2}}^2 (x_1)^{2k_1} (x_2)^{2k_2} \Eb \left [|y|^{2 k_3}\right]\\
    &=\sum_{k_3=0}^{m} {\binom{m}{k_3, m-k_3}}^2 \Eb \left [|y|^{2 k_3}\right] \sum_{k_1+ k_2=m-k_3} {\binom{m-k_3}{k_1, k_2}}^2 (x_1)^{2k_1} (x_2)^{2k_2} .
\end{align*}
Based on Corollary \ref{lem leb poly}, the value of
 $\sum_{k_1+ k_2=m-k_3} {\binom{m-k_3}{k_1, k_2}}^2 (x_1)^{2k_1} (x_2)^{2k_2}$ is maximized when $x_1^2=x_2^2 = \frac{1-t}{2}$.  Because we assume the positiveness, we will have that $x_1=x_2 = \sqrt{\frac{1-t}{2}}$. This concludes the proof. \qed

\subsection{Proof of \cref{lem: vector case c}}
Without loss of generality, we assume $c_1 \neq c_2$ while $c_1^2+c_2^2$ is a constant. Similar to the proof in \cref{se: ap B2}, it is sufficient to prove that making $c_1=c_2$ can increase the value in \cref{eq:lem vector case c}. Then, the general result in \cref{lem: vector case c} can be obtained by induction.
First of all, separating $c_1, c_2$ with $c_m,\forall m\ge 3$ gives the following,
\begin{align*}
&\sum_{k_1+\ldots+k_M=t} {\binom{t}{k_1, \ldots,k_M}}^2    \prod_{m=1}^{M}c_m^{2k_m}\\
 &= \sum_{k_1+\ldots+k_M=t} {\binom{t}{k_1, \ldots,k_M}}^2    c_1^{2k_1}c_2^{2k_2}\prod_{m=3}^{M}c_m^{2k_m}\\
 &=\sum_{t_0=0}^{t} \sum_{k_1+k_2=t_0} \sum_{k_3+,\ldots,k_M=t-t_0}{\binom{t}{k_1, \ldots,k_M}}^2    c_1^{2k_1}c_2^{2k_2}\prod_{m=3}^{M}c_m^{2k_m} \\
  &=\sum_{t_0=0}^{t} \sum_{k_1+k_2=t_0} \sum_{k_3+,\ldots,k_M=t-t_0}
  {\binom{t}{t_0, t-t_0}}^2{\binom{t_0}{k_1,k_2}}^2 {\binom{t-t_0}{k_3,\ldots,k_m}}^2 c_1^{2k_1}c_2^{2k_2}\prod_{m=3}^{M}c_m^{2k_m}.
\end{align*}

Then, according to \cref{lem leb poly}, we have the following inquality,
\begin{align*}
  \sum_{k_1+k_2=t_0}  {\binom{t_0}{k_1, k_2}}^2      c_1^{2k_1}c_2^{2k_2} \le \sum_{k_1+k_2=t_0}  {\binom{t_0}{k_1, k_2}}^2     \left(\frac{c_1^2+c_2^2}{2}\right)^{t_0},
\end{align*}
where the equality holds when $c_1=c_2=\sqrt{\frac{c_1^2+c_2^2}{2}}$. This concludes the proof.

\subsection{Proof of \cref{lem:holder N2}}

We mainly use the Hölder's inequality for the proof of \cref{lem:holder N2}. Specifically, for positive random variables $X$ and $Y$, if $1/p+1/q=1$, then Hölder's inequality shows that
\begin{align*}
\Eb[X Y] \le (\Eb[X^p])^{1/p} (\Eb[Y^q])^{1/q}.
\end{align*}
For the posted problem, we let $X=X_1^{k_1}$ and $Y=X_2^{k_2}$, then we have
\begin{align*}
\Eb[X_1^{k_1} X_2^{k_2}] \le (\Eb[X_1^{k_1 p}])^{1/p} (\Eb[X_2^{k_2 q}])^{1/q}.
\end{align*}
which concludes the proof of \cref{eq:holder N2 pq}.

When $k_1=0$ or $k_2=0$, the inequality in \cref{eq:holder N2 max} holds trivially. Here, without loss of generality, we assume $k_1,k_2 \neq 0$.
By letting $p=\frac{k_1+k_2}{k_1}$ and $q=\frac{k_1+k_2}{k_2}$ in the expression \cref{eq:holder N2 pq}, one can check  that $1/p+1/q=1$. Then, the following inequality holds due to the Hölder's inequality,
\begin{align*}
\Eb[X_1^{k_1} X_2^{k_2}] &\le (\Eb[X_1^{k_1+k_2}])^{\frac{k_1}{k_1+k_2}} (\Eb[X_2^{k_1+k_2 }])^{\frac{k_2}{k_1+k_2}}\\
&\le \max\{ \Eb[X_1^{k_1+k_2}], \Eb[X_2^{k_1+k_2}] \}.
\end{align*}
Thus, this concludes the proof of  the result in \cref{eq:holder N2 max}. \qed

\subsection{Proof of \cref{lem:holder N general}}
  Without loss of generality, we assume $k_n >0, \forall n$. Using the Hölder's inequality, we have
  \begin{align} \label{eq:supp 1}
\Eb\left[X_1^{k_1} \cdots X_{N-1}^{k_{N-1}}X_{N}^{k_{N}}\right]\le \left(\Eb\left[\left(X_1^{k_1} \cdots X_{N-1}^{k_{N-1}}\right)^{\frac{t}{t-k_N}}~\right] \right)^{\frac{t-k_N}{t}}\left(\Eb\left[(X_{N}^{k_{N}} )^ {t/k_N}\right]\right)^{k_N/t},
\end{align}
where $p=t/(t-k_N)$ and $q=t/k_N$. Simplifying the \gls{RHS} in \cref{eq:supp 1} yields
\begin{align} \label{eq: sub N}
\Eb\left[X_1^{k_1} \cdots X_{N-1}^{k_{N-1}}X_{N}^{k_{N}}\right] \le \left(\Eb\left[\left(X_1^{k_1} \cdots X_{N-1}^{k_{N-1}}\right)^{t/(t-k_N)}~\right] \right)^{(t-k_N)/t}\left(\Eb\left[X_{N}^{t}\right]\right)^{k_N/t}.
\end{align}
Again, using Hölder's inequality for the first part of \gls{RHS} of \cref{eq: sub N} with $p=\frac{t-k_N}{t-k_N-k_{N-1}}$ and $q={\frac{t-k_N}{k_{N-1}}}$, we have
\begin{align*}
&\Eb\left[\left(X_1^{k_1} \cdots X_{N-1}^{k_{N-1}}\right)^{t/(t-k_N)}\right] \\
&= \Eb\left[\left(X_1^{k_1} \cdots X_{N-2}^{k_{N-2}}\right)^{t/(t-k_N)}  X_{N-1}^{k_{N-1}\frac{t}{t-k_N}}\right]\\
&\le  \left( \Eb\left[\left(X_1^{k_1} \cdots X_{N-2}^{k_{N-2}}\right)^{\frac{t}{t-k_N}\frac{t-k_N}{t-k_N-k_{N-1}}}  \right] \right)^{\frac{t-k_N-k_{N-1}}{t-k_N}} \left( \Eb \left[ \left(X_{N-1}^{k_{N-1}\frac{t}{t-k_N}} \right)^{\frac{t-k_N}{k_{N-1}}}\right]\right)^{\frac{k_{N-1}}{t-k_N}} \\
& = \left( \Eb\left[\left(X_1^{k_1} \cdots X_{N-2}^{k_{N-2}}\right)^{\frac{t}{t-k_N-k_{N-1}}}  \right]  \right)^{\frac{t-k_N-k_{N-1}}{t-k_N}} \left( \Eb \left[ X_{N-1}^t\right]\right)^{\frac{k_{N-1}}{t-k_N}}.
\end{align*}
Substituting the above equation into \cref{eq: sub N} gives
\begin{align}
&\Eb\left[X_1^{k_1} \cdots X_{N-1}^{k_{N-1}}X_{N}^{k_{N}}\right] \nonumber  \\
&\le \underbrace{ \left(\Eb\left[\left(X_1^{k_1} \cdots X_{N-2}^{k_{N-2}}\right)^{\frac{t}{t-k_N-k_{N-1}}}  \right]  \right)^{\frac{t-k_N-k_{N-1}}{t-k_N}}}_{\text{first part}} \underbrace{ \left( \Eb \left[ X_{N-1}^t\right]\right)^{\frac{k_{N-1}}{t}} \left(\Eb\left[X_{N}^{t}\right]\right)^{\frac{k_N}{t}}}_{\text{remaining parts}}. \label{eq:hol se}
\end{align}
For the first part in  \cref{eq:hol se}, we  iteratively utilize the Hölder's inequality, and  finally obtain the following,
\begin{align*}
\Eb\left[\prod_{n=1}^{N} X_n^{k_n}\right] &\le \prod_{n=1}^N\left(\Eb\left[X_{n}^{t}\right]\right)^{k_n/t}\\
& \le \max_n \Eb\left[X_n^t\right].
\end{align*}
This concludes the proof of \cref{lem:holder N general}.
\qed

\section{\uppercase{Additional Experiments}} \label{app:C}
\begin{figure}[t]
\centering  
\includegraphics[width=0.55\textwidth]{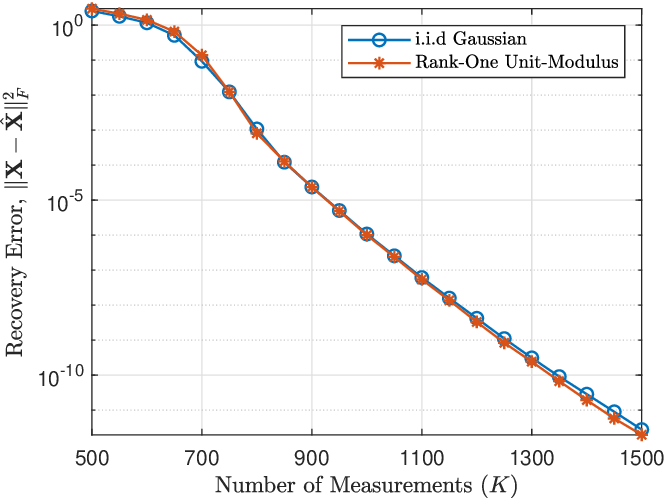}
\caption{Recovery Error of Alternating Minimization Method by Using Rank-One Unit-Modulus and i.i.d. Gaussian Measurements with Different Number of Measurements} 
\label{fig:recover AM}   
\end{figure}

\begin{figure}[t]
\centering  
\includegraphics[width=0.55\textwidth]{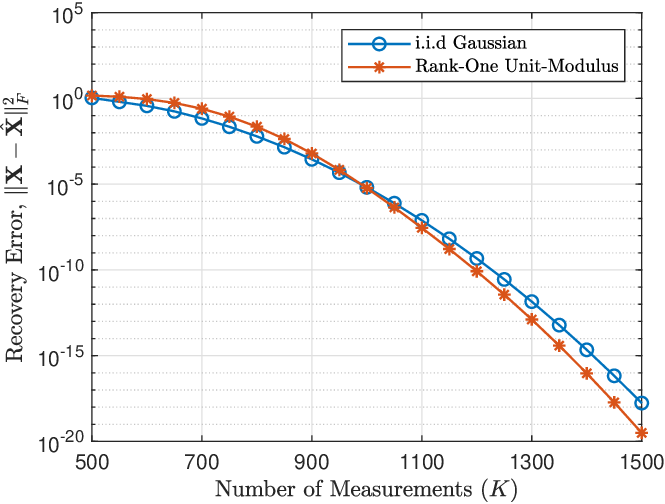}
\caption{Recovery Error of Gradient-Based Method by Using  Rank-One Unit-Modulus and i.i.d. Gaussian Measurements with Different Number of Measurements} 
\label{fig:recover GD}   
\end{figure}

In this section, we evaluate the recovery performance of non-convex low-rank matrix recovery algorithms by using the  rank-one  unit-modulus
measurements and \gls{iid} Gaussian measurements. Similar to the settings in \cref{fig:recover different K}, we randomly generate the target  low-rank matrix $\bX\in \C^{40\times 80}$ with a rank of $r=5$, and the number of measurements  $K$ ranges from $500$ to $1500$. There are two typical non-convex algorithms considered in the experiments. Specifically, in \cref{fig:recover AM} , we utilize the alternating minimization method \citep{jain2013low} to recover the low-rank matrix from the   rank-one  unit-modulus
measurements or the \gls{iid} Gaussian measurements. 
In \cref{fig:recover GD}, we employ the gradient-based method \citep{chi2019nonconvex} for matrix recovery task. 

As we can see in  \cref{fig:recover AM,fig:recover GD},  for both typical non-convex methods, i.e., alternating minimization method and gradient-based method, the designed rank-one measurements with random unit-modulus vectors achieves a recovery performance similar to that of  \gls{iid} Gaussian measurements. 
Therefore, by integrating the results of experiment in \cref{fig:recover different K}, regardless of whether convex or non-convex optimization algorithms are utilized,  the rank-one unit-modulus measurements always exhibit a similar matrix  recovery performance to that of \gls{iid} measurements, which is  attributed to the proven RIP results.

\medskip

\vfill
\end{document}